\documentclass[a4paper,twoside]{article}
\usepackage[usenames,dvipsnames,svgnames,table]{xcolor}
\usepackage[utf8]{inputenc}
\usepackage[T1]{fontenc}
\usepackage{graphicx}
\usepackage{grffile}
\usepackage{longtable}
\usepackage{wrapfig}
\usepackage{rotating}
\usepackage[normalem]{ulem}
\usepackage{amsmath}
\usepackage{textcomp}
\usepackage{amssymb}
\usepackage{capt-of}
\usepackage{hyperref}
\usepackage[margin=1in]{geometry}
\usepackage{authblk}
\usepackage{subcaption}
\usepackage{float}
\usepackage{wrapfig}
\restylefloat{table}
\newcommand{\matcont}{\textsc{MatCont}}
\newcommand{\matcontm}{\textsc{MatContM}}
\newcommand{\clmatcont}{\textsc{Cl\_MatCont}}

\author[1]{Mark Blyth}
\author[2]{Ludovic Renson \thanks{Co-last and co-corresponding authors: l.renson@imperial.ac.uk; lucia.marucci@bristol.ac.uk}}
\author[1,3,4]{Lucia Marucci $^\ast$}
\affil[1]{Department of Engineering Mathematics, University of Bristol, Bristol BS8 1UB, UK.}
\affil[2]{Department of Mechanical Engineering, Imperial College London, South Kensington Campus, London SW7 2AZ, UK}
\affil[3]{School of Cellular and Molecular Medicine, University of Bristol, Bristol BS8 1TD, UK.}
\affil[4]{BrisSynBio, Bristol BS8 1TQ, UK}
\date{}
\title{Tutorial of numerical continuation and bifurcation theory for systems and synthetic biology}
\hypersetup{
 pdfauthor={},
 pdftitle={Tutorial of numerical continuation and bifurcation theory for systems and synthetic biology},
 pdfkeywords={},
 pdfsubject={},
 pdfcreator={Emacs 26.3 (Org mode 9.1.9)}, 
 pdflang={English}}
\begin{document}

\maketitle
\begin{abstract}
Mathematical modelling allows us to concisely describe fundamental principles in biology.
Analysis of models can help to both explain known phenomena, and predict the existence of new, unseen behaviours.
Model analysis is often a complex task, such that we have little choice but to approach the problem with computational methods.
Numerical continuation is a computational method for analysing the dynamics of nonlinear models by algorithmically detecting bifurcations.
Here we aim to promote the use of numerical continuation tools by providing an introduction to nonlinear dynamics and numerical bifurcation analysis.
Many numerical continuation packages are available, covering a wide range of system classes; a review of these packages is provided, to help both new and experienced practitioners in choosing the appropriate software tools for their needs.
\end{abstract}

\section{Introduction}
\label{sec:org1800635}
Computational biology uses mathematical tools to understand the mechanisms that orchestrate life and living processes \cite{beard2005computational}.
Differential equations are a powerful tool for describing how biological processes behave, based on the values and rate of change of a set of variables.
Both the equations and their solutions can be analysed to explain observed phenomena, and predict novel and unseen behaviours.

A differential equation is linear only if its rates of change follow simple proportional relationships.
Biological systems are rarely linear, and instead fall into the wide category of nonlinear systems \cite{beuter2003nonlinear}.
Nonlinearity arises from complex interplays between the constituent components of a system, such as the subtle mutual dependencies of a neuron's ionic currents \cite{izhikevich2007dynamical}, or the complex feedback loops of mammalian erythropoiesis \cite{belair1995age}.
Nonlinearity allows for a rich set of non-trivial behaviours, at the expense of simplicity.
Nonlinear systems are rarely analytically tractable -- it is not generally possible to obtain a closed-form solution to a nonlinear differential equation.
As a result, the field of nonlinear dynamics considers how best to analyse the equations without actually solving them.

The challenges in working with nonlinear systems are justified by their immense explanatory power.
Classic examples include the work of Hodgkin and Huxley \cite{hodgkin1952quantitative}, which laid the foundations of modern neuroscience; the Mackey-Glass equation \cite{mackey1977oscillation}, which describes the nuanced effects of time-delayed feedback on respiratory and hematopoietic diseases; and the Lotka-Volterra model \cite{volterra1928variations}, which describes the interactions between populations of predators and prey.
Synthetic biology makes extensive use of nonlinear models and their associated analysis tools, to design and represent artificially engineered gene networks \cite{di2012predicting,kulkarni2014systems}.

Numerical continuation is a standard method within the nonlinear dynamics community.
In this context, it is used to computationally analyse nonlinear differential equations.
Here we aim to introduce the method to those without a background in nonlinear dynamics. 
Examples are provided to highlight existing biological applications, and to demonstrate the applicability of nonlinear dynamics to biology.
Section \ref{sec:orgbb05e5e} introduces some key ideas from nonlinear dynamics.
Section \ref{sec:org29b9f03} discusses a nonlinear model, which is then used in a numerical continuation experiment in section \ref{sec:orgd97deef}.
This tutorial helps to motivate some of the core ideas from the field.
A wide range of software packages are available for numerical continuation, which are reviewed in section \ref{sec:orga594d2d}.
Note that while some existing publications include a survey of these tools (see \cite{meijer2009numerical,govaerts2007interactive}) they focus primarily on the history and design approaches of the software; we instead consider the various practicalities and usage cases, to help users choose the appropriate tool for their needs.

\section{Nonlinear dynamics}
\label{sec:orgbb05e5e}
Here we introduce some foundational ideas from nonlinear dynamics, and motivate the concept of a bifurcation.
These ideas are then demonstrated in practice in section \ref{sec:orgd97deef}.
See \cite{strogatz2018nonlinear,guckenheimer2013nonlinear,kuznetsov2013elements} for a comprehensive discussion of bifurcations and nonlinear dynamics, and \cite{beuter2003nonlinear,izhikevich2007dynamical,hoppensteadt2012weakly} for an introduction from a mathematical biology perspective.
 The authors particularly recommend \cite{strogatz2018nonlinear} for a very readable entry-point to the field.

Differential equations describe how the state of a system changes with respect to an independent variable.
Time is often taken as the independent variable, in which case the differential equations describe the temporal evolution of states.
A system state quantifies all necessary information required to describe and predict the behaviours of interest.
For some classes of differential equation, states must be functions, such as the population density of a species in a partial differential equation model of predator-prey interaction, or the previous values of red blood cell counts in a delay differential equation for erythropoiesis; see \cite{murray2001mathematical} for a wide variety of examples.
Such classes of differential equation are not considered here.
Instead, the proceeding work considers ordinary differential equations, where a set of dependent variables evolves under a single time-like dependent variable, with the evolution depending only on the current value of the state.

The qualitative behaviour of a system is referred to as its dynamics.
For example, the dynamics could be quiescent if the system remains settled at a steady-state, or oscillatory if the state fluctuates in a consistent, repetitive manner, or chaotic if the state changes deterministically but unpredictably.
A system can settle to different dynamics from different initial states, for example when quiescent and oscillatory behaviours coexist.

\begin{figure}
    \center
    \includegraphics[width=.8\textwidth]{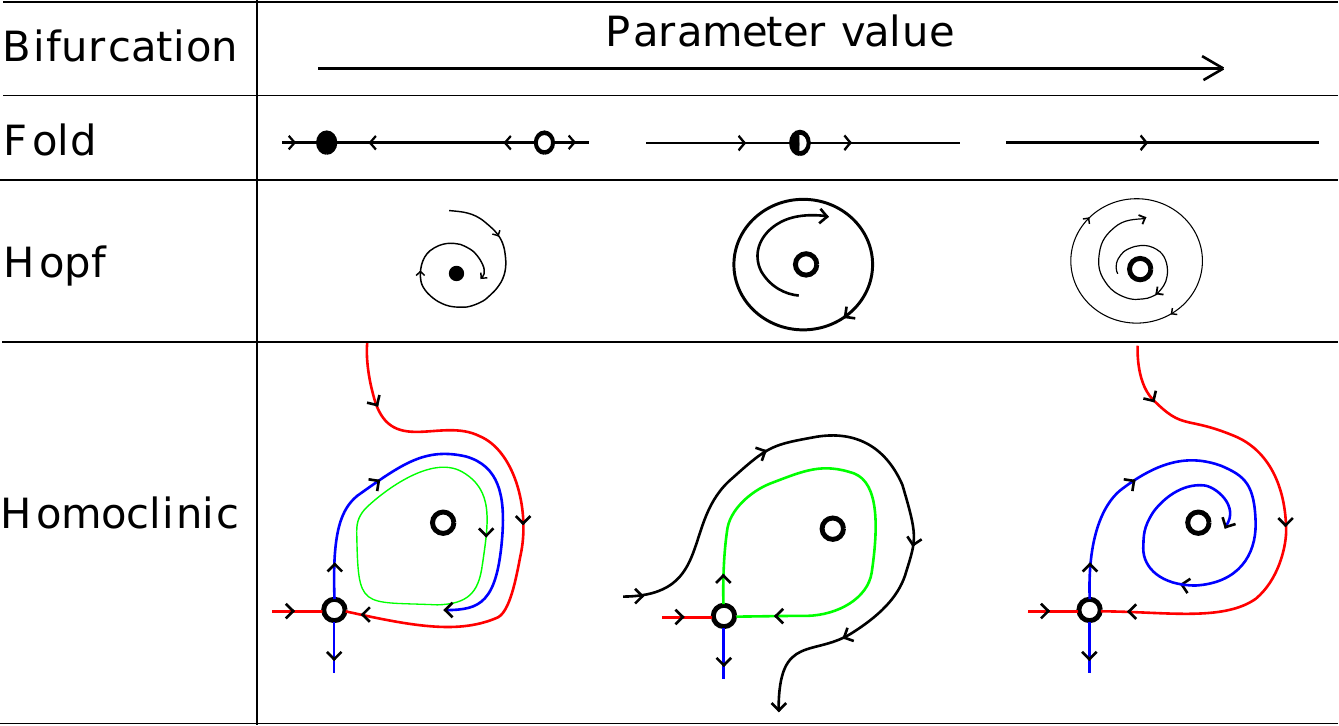}
    \caption{Sketch phase portraits for a fold, Hopf, and homoclinic bifurcation. Solid circles represent stable equilibria, and hollow circles represent unstable equilibria. Arrows represent direction of flow. In the fold bifurcation, two equilibria coalesce and annihilate at the bifurcation point. In the subcritical Hopf bifurcation, the equilibrium loses stability at the bifurcation point, and the flow is pushed onto a spawned limit cycle. Immediately before a homoclinic bifurcation, the unstable manifold of the saddle (blue) is sandwiched between a limit cycle (green) and the stable manifold (red); at the bifurcation point, these three objects meet and form a homoclinic connection, annihilating the limit cycle.}
    \label{fig:bifs}
\end{figure}

System dynamics generally depend a set of external parameters.
The parameter dependence is often non-trivial, such that changing a parameter value by just a small amount can sometimes lead to drastic changes in the system behaviours.
An extreme example of this can be seen in the canard explosion of a food-chain model \cite{deng2004food}.
Here, the population size transitions between steady and rapidly oscillating in an exponentially small region of parameter space.
When a change the parameter value causes a qualitative change in dynamics, a bifurcation is said to have occurred.
A bifurcation occurs at a parameter value if a system contains dynamics \emph{within a neighbourhood} of the parameter that differ from the dynamics \emph{at} the parameter.
That is, no matter how small a distance we move away from the bifurcation point, we can always find system behaviours that are qualitatively different to those that occur at the bifurcation point.
A bifurcation diagram shows how invariant sets (such as steady-state position) or measures on the system (such as oscillation amplitude) change, as a parameter is varied.
The simplest bifurcations include the fold, Hopf, and homoclinic bifurcations, as sketched in figure \ref{fig:bifs}.
These can be observed when only a single parameter is varied.

A fold bifurcation occurs when a stable and unstable equilibrium annihilate.
Pairs of fold bifurcations often lead to switch-like behaviours in biological systems \cite{ozbudak2004multistability}.
This happens when a region of bistability occurs, where two stable equilibria coexist.
The switching behaviour arises when an external stimulus pushes the system from one equilibrium to the other.
Fold-induced multistability can model a wide range of behaviours, including metabolic pathway dynamics \cite{diaz2010bistable}, visual perception tasks \cite{chialvo1993modulated}, many biological feedback systems \cite{angeli2004detection}, and cell specialisation, whereby cells specialise by losing unspecialised states through fold bifurcations \cite{ferrell2012bistability,marucci2017nanog}.
Bistable dynamics are regularly exploited in synthetic biology, to design and predict system dynamics \cite{zheng2010mathematical}.
For example, saddle-node bifurcations in eukaryotic cell signalling pathways leads to bistability \cite{wang2006bistability}, which has been used to create a cellular hysteresis-loop oscillator \cite{chickarmane2007oscillatory}.

A homoclinic bifurcation happens when a limit cycle approaches, collides with, and is annihilated by a saddle equilibrium.
This is sketched, in the case of a stable limit cycle, in figure \ref{fig:bifs}.
Homoclinic bifurcations are a key cause of spike termination in neurons \cite{izhikevich2007dynamical}.

Hopf bifurcations are the simplest way for a system to transition to oscillatory behaviour.
In a Hopf bifurcation, an equilibrium changes stability and a limit cycle appears or disappears around it, as sketched in figure \ref{fig:bifs}.
This can occur, for example, in a predator-prey system -- the system undergoes a Hopf bifurcation when feedbacks between the competing populations cause oscillations to emerge in the population sizes \cite{fussmann2000crossing}.
Hopf bifurcations can be used when designing synthetic biological systems.
A typical use is determining the regions in parameter space that allow for some desired dynamics.
For example, a synthetic gene-metabolic system has been designed to display oscillatory behaviours; the glycolytic flux required to induce oscillations is predicted using bifurcation analysis \cite{fung2005synthetic}.

The detection and analysis of bifurcations is often challenging.
A rigorous method is to study the equations analytically, and determine where in the parameter space a bifurcation occurs.
Textbooks such as \cite{strogatz2018nonlinear,guckenheimer2013nonlinear} discuss this approach, with \cite{kuznetsov2013elements} providing a comprehensive overview.
Analytical calculations are challenging for complex systems; as a result, computer-based methods are employed instead.
One possible approach is brute force, whereby one simulates a system for a variety of parameter values, noting the resultant behaviours.
This method works for simple systems with few parameters, however it is impractical for studying the complex bifurcation structures typically observed in biology.
Numerical continuation is the standard approach to computational bifurcation analysis.
It is more efficient and rigorous than brute force, and significantly easier than analytical methods.
Numerical continuation is discussed and demonstrated on a model in section \ref{sec:orgd97deef}.

\section{Case study: bursting neuron models}
\label{sec:org29b9f03}
Here a nonlinear model is introduced.
This model will be analysed using numerical continuation in section \ref{sec:orgd97deef}.
The Hindmarsh-Rose model \cite{hindmarsh1984model} is chosen for its ability to display rich non-trivial dynamics, whilst also allowing a simple planar analysis.
The model describes bursting dynamics in neurons.
Bursting is characterised by a periodic transition between an active spiking state and a quiescent rest state.
The cell membrane potential oscillates rapidly during spiking, and remains relatively constant during the quiescent phase.

The Hindmarsh-Rose model is given \cite{hindmarsh1984model} by
\begin{align}
\frac{\mathrm{d} x}{\mathrm{d} t} &= y - ax^3 +bx^2 -z + I~,\\ \nonumber
\frac{\mathrm{d} y}{\mathrm{d} t} &= c- dx^2 -y~,\\ 
\frac{\mathrm{d} z}{\mathrm{d} t} &= r\left[s(x-x_R)-z\right]~,\nonumber
\end{align}
where \(a=1,b=3,c=1,d=5,s=4,x_R=-1.6, r=0.001\) are dimensionless parameters \cite{hindmarsh1984model}.
Variables \((x,y,z)\) form a non-dimensional state vector.
The individual state variables describe some aspect of the system that we are interested in learning about: \(x\) models the membrane potential of a bursting cell, \(y\) the main currents into and out of the cell, and \(z\) an adaptive (calcium-like) current.
The parameters describe the set of controllable and uncontrollable inputs to the system.
For example, parameter \(I\) models an externally applied current, and is easily controllable by an experimenter; conversely, \(r\) measures the timescale separation of the adaptive current, which is less trivial to modify experimentally.

The magnitude of the \(z\)-variable's time-derivative is approximately \(r\) times smaller than that of the \(x\) and \(y\) variables.
The \(x,y\) variables thus change much more rapidly than \(z\), and are said to form the `fast subsystem'; \(z\), changing slowly, forms the slow subsystem.
Consider the singular limit \(r\to0\); the slow subsystem ceases to change and \(z\) becomes constant, fixed at its initial condition \(z(0)\).
Our investigative procedure, as pioneered by Rinzel \cite{rinzel1985bursting}, exploits this as follows.
The slow variable \(z\) is treated as a parameter of the fast subsystem; the dependence of the fast-subsystem dynamics on \(z\) is investigated with numerical continuation; the slow subsystem is then reintroducted, revealling the cause of bursting dynamics.

\section{Numerical continuation}
\label{sec:orgd97deef}
\begin{figure}
    \center
    \includegraphics[width=\textwidth]{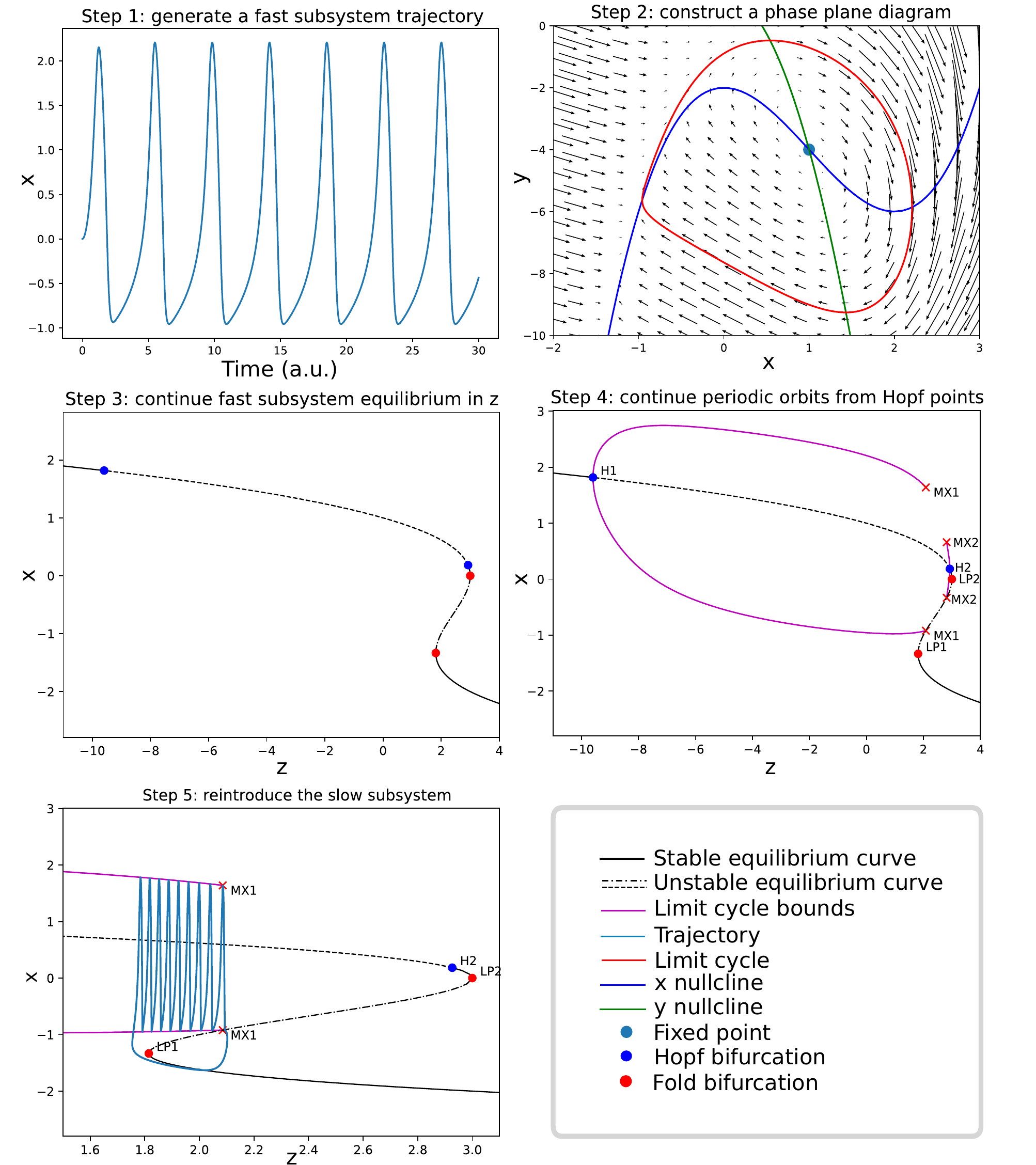}
    \caption{Investigating the cause of bursting in the Hindmarsh Rose model. Dark blue dots, labelled `H', indicate a Hopf bifurcation. Red dots, labelled `LP', indicate a limit point, or fold, bifurcation. Red crosses, labelled `MX', indicate the end of continuation, here demarcating a homoclinic bifurcation. A solid black line represents a stable equilibrium location; a dashed black line indicates an unstable equilibrium location; a purple line represents the maximum and minimum values of a limit cycle. Trajectories are shown as light blue lines.}
    \label{fig:tutorial}
\end{figure}

Here we demonstrate a bifurcation analysis on the Hindmarsh-Rose model.
Figure \ref{fig:tutorial} shows a diagram of all the steps followed here.
The analysis is performed using PyDSTool \cite{clewley2012hybrid}, however the same results could be achieved using other continuation tools; see section \ref{sec:orga594d2d} for a survey.

To promote spike generation in the neuron model, a positive current-like perturbation \(I\) is applied; \(I=2\) is chosen arbitrarily.
First, the fast-subsystem is analysed.
As \(z\) is being treated as a parameter, we must also choose an initial value for it; \(z=0\) is chosen arbitrarily.

\uline{\textbf{Step 1:}}
Initially, we have no knowledge of the system dynamics; therefore, we first try to learn how the fast-subsystem behaves at standard parameter values.
This is done by numerically integrating the fast subsystem differential equations from some initial state \((x(0), y(0))\).
In doing so, we gain some insights into the normal behaviours of the system, which will be useful for the next steps where we look at changes in its behaviour.

In figure \ref{fig:tutorial} panel one we plot a time-series of the \(x\) variable, and see that the system settles to sustained oscillations.
This necessitates a stable limit cycle in the fast subsystem -- an attracting, closed curve along which states are visited and re-visited in a repetitive manner.

\uline{\textbf{Step 2:}}
A rest-state (equilibrium) must exist inside this limit cycle; such a rest-state provides a starting point for numerical continuation, so step 2 is to locate it.
In figure \ref{fig:tutorial} panel two, a phase plane diagram is created.
Phase plane analysis is a graphical method for showing the dynamics of a low-dimensional nonlinear system.
A point \((x,y)\) on the phase plane represents a system state; arrows indicate the flow direction -- the direction a state moves in when acted upon by the differential equations; the size of the arrows indicate how fast a state moves at any given point.
Trajectories map the locuses of states; they are plotted as lines on a phase plane.
A single, closed trajectory is shown; this is the limit cycle that produces oscillating dynamics.
Nullclines show the sets of points in the state space where a coordinate has a zero time-derivative.
They separate the phase plane into regions of same-direction flow.
At the intersection of the nullclines, the time-derivative is zero in all directions and the state cannot move in time; hence, the system is at an equilibrium (a rest state).
As predicted, the phase plane shows there is an equilibrium inside the limit cycle.

\uline{\textbf{Step 3:}}
For step three, we use continuation to change \(z\) and track the corresponding changes in the equilibrium position.
The resulting plot, shown in figure \ref{fig:tutorial} panel 3, is a bifurcation diagram.
The black lines, solid and dotted, show the \(x\) location of the equilibria.
As this curve shows the location of equilibrium-points, it is often referred to as an equilibrium-point curve.
Equilibria can be stable or unstable: if nearby states move away from an equilibrium, it is unstable; otherwise, it is stable.
Stable equilibria are marked with a solid line on an equilibrium-point curve.
Unstable equilibria are marked with a dotted line.

Continuation algorithms implement test functions, which are used to detect bifurcations along an equilibrium-point curve.
Four bifurcations are identified -- Hopf points H1 and H2, and fold points LP1 and LP2.
While the equilibrium-point curve changes between solid and dotted lines at points LP1 and LP2, no stability change occurs.
Instead, a stable and unstable equilibrium meet, coalesce, and annihilate in a fold bifurcation.
Hopf bifurcations occur at points H1 and H2 in figure \ref{fig:tutorial} panel 3.

\uline{\textbf{Step 4:}}
In figure \ref{fig:tutorial} panel four, numerical continuation is used again, this time to track the limit cycles arising from the Hopf bifurcations.
Two families of limit cycles are found -- one from each of the two Hopf bifurcations.
The limit cycle family from Hopf H1 produces the main oscillatory behaviours of the Hindmarsh-Rose model.
The maximum and minimum values of these limit cycles are plotted as purple lines on the bifurcation diagram.
Homoclinic bifurcations are observed at points MX1 and MX2 in figure \ref{fig:tutorial}.
Note that MX does not mean a homoclinic bifurcation has been detected, but that the continuation algorithm failed to converge.
It is up to the user to identify that the convergence failure is due to the limit cycles disappearing in a homoclinic bifucation.

\uline{\textbf{Step 5:}}
In figure \ref{fig:tutorial} panel five, we reintroduce the slow subsystem to uncover the cause of bursting.
A trajectory of the full system is computed and plotted in the \((z,x)\) plane, on top of the bifurcation diagram.
Bursting behaviour is seen to arise as follows.
Bistability exists between an equilibrium and a limit cycle.
This bistable region starts at a fold bifurcation on the left, and ends at a homoclinic bifurcation on the right, respectively labelled LP1 and MX1 in figure \ref{fig:tutorial}.
The slow subsystem periodically drives the fast subsystem across this pair of bifurcations.
The state sits at an equilibrium until it disappears though fold bifurcation LP1.
This causes the system to jump onto the limit cycle originating at Hopf H1, with the limit cycle producing spiking dynamics.
Now \(z\) flows in the opposite direction, until the limit cycle disappears through homoclinic bifurcation MX1.
The system jumps back to the equilibrium, and spiking is terminated.
Once again the flow of \(z\) changes direction, and the pattern repeats.
This periodic transition between an equilibrium and limit cycle causes a transition between quiescent and spiking states resepectively, and hence bursting behaviours arise.
The use of numerical continuation allows us to find the points at which these bifurcations occur, and to generate diagrams which graphically explain the behaviour.
With experience, these diagrams become an invaluable tool to understanding the dynamics of nonlinear systems.

\section{Review of continuation software}
\label{sec:orga594d2d}
Numerous software packages have been written for performing numerical continuation.
Here, \matcont{} \cite{dhooge2008new}, PyDSTool \cite{clewley2012hybrid}, XPPAUTO \cite{ermentrout2002simulating}, and CoCo \cite{dankowicz2013recipes} -- the four most common continuation tools -- are discussed.
These are all applicable to ordinary differential equations.
Tools for other system classes are surveyed in section \ref{sec:orgf65dff9}.

Tables \ref{tab:orgf0e37d4} and \ref{tab:org80b7563} provide a comparison of key features of the main continuation software tools.
Table \ref{tab:orgf0e37d4} shows a summary of the general software features; table \ref{tab:org80b7563} shows which of various bifurcations each software tool is able to detect, and which it is able to continue.
Each tool has its own best-usage criteria.
For example, \matcont{} is able to detect and continue the most bifurcations of the tools, however it can only simulate ordinary differential equations.
XPPAUTO can simulate a wider range of system types, at the expense of detecting fewer bifurcation.
CoCo is a development toolbox, so can be applied to any continuation problem.
PyDSTool and XPPAUTO both use low-level integrators and continuation algorithms, making them significantly faster than CoCo and \matcont{}.
The choice of tool therefore comes down to user preferences such as coding language and interface type, and performance factors such as speed and capabilities.
Note that some bifurcations which are not detected by a software tool can still be identified by users.
For example, XPPAUTO cannot explicitly detect cusp bifurcations, however they are easily spotted from the intersection of two fold manifolds.

\begin{table}[]
\caption{\label{tab:orgf0e37d4}
Comparison table of systems each tool can either simulate or perform continuation on, as well as a selection of other relevant tool features. Notes: (1) the capabilities of the entire Computational Continuation Core are considered here. As discussed in section \ref{sec:orgb7cb7d5}, one can develop custom routines in CoCo to study any class of system to which continuation can be applied. Nevertheless, not all systems will be covered by the bundled \emph{ep} and \emph{po} toolboxes, so users may have to develop their own CoCo interfaces. (2) \matcontm{} is a version of \matcont{} built for studying maps; (3) \clmatcont{} is an extensible script-based alternative to GUI \matcont{}; (4) creative Commons Attribution-NonCommercial-ShareAlike. Allows for modification and redistribution, subject to constraints; (5) a MATLAB license is required to run the software; (6) while XPPAUTO has a simulation environment, it is much less closely integrated with the continuation environment than that of PyDSTool and \matcont{}.}
\centering
\begin{tabular}{lllll}
\hline
Feature & \matcont{} & CoCo\(^{\text{(1)}}\) & XPPAUTO & PyDSTool\\
\hline
ODEs & y & y & y & y\\
PDEs (discretized) & n & y & y & n\\
DDEs & n & y & y & limited\\
SDEs & n & n & y & limited\\
DAEs & n & y & y & y\\
BVPs & n & y & y & n\\
Maps & n\(^{\text{(2)}}\) & y & y & y\\
Hybrid systems & n & y & limited & y\\
Coding interface & n\(^{\text{(3)}}\) & y & n & y\\
Main language & MATLAB & MATLAB & C & Python\\
License & CC BY-NC-SA\(^{\text{(4,5)}}\) & Unspecified\(^{\text{(5)}}\) & GNU GPL v2 & BSD 3-clause\\
Simulation tools & y & n & y\(^{\text{(6)}}\) & y\\
Extensible & n\(^{\text{(3)}}\) & y & n & y\\
\end{tabular}
\end{table}

\begin{table}[]
\caption{\label{tab:org80b7563}
Comparison of points and bifurcations each software tool can handle. \emph{D} means a tool can detect a given point; \emph{C} means a tool can perform numerical continuation of a given point. Notes: (1) only the capabilities of the bundled \emph{ep} and \emph{po} toolboxes are considered here; (2) custom routines can easily be developed to study other types of bifurcation.}
\centering
\begin{tabular}{lrllll}
\hline
Bifurcation Type & Codimension & \matcont{} & CoCo\(^{\text{(1,2)}}\) & XPPAUTO & PyDSTool\(^{\text{(2)}}\)\\
\hline
Equilibrium & 0 & C & C & D,C & D,C\\
Limit cycle & 0 & C & C & C & C\\
Limit point & 1 & D,C & D,C & D,C & D,C\\
Hopf & 1 & D,C & D,C & D,C & D,C\\
Limit point of cycles & 1 & D,C & D,C & - & D\\
Neimark-Sacker & 1 & D,C & D,C & D,C & D,C\\
Period doubling & 1 & D & D,C & D,C & D,C\\
Homoclinic & 1 & C & - & C & -\\
Cusp & 2 & D & - & - & D\\
Bogdanov Takens & 2 & D & D & - & D\\
Zero-Hopf & 2 & D & - & - & D\\
Double Hopf & 2 & D & - & - & D\\
Generalised Hopf & 2 & D & - & - & D\\
Cusp point of cycles & 2 & D & - & - & -\\
Chenciner & 2 & D & - & - & -\\
Fold-Neimark-Sacker & 2 & D & - & - & -\\
Flip-Neimark-Sacker & 2 & D & - & - & -\\
Fold-Flip & 2 & D & - & - & -\\
Double Niemark-Sacker & 2 & D & - & - & -\\
Generalised flip & 2 & D & - & - & -\\
\end{tabular}
\end{table}

\subsection{XPPAUTO}
\label{sec:org53446f0}

XPPAUTO (also called XPPAUT, XPP) is a combined simulation and analysis package \cite{ermentrout2002simulating}.
It provides an interface to AUTO \cite{doedel1981auto,doedel1997auto,doedel2007auto} for numerical continuation.
XPPAUTO is one of the oldest dynamical systems tools to still see regular use.
A large community and a range of resources are available, see for example \cite{ermentrout2002simulating}.
Nevertheless, the age of the software also lends itself to a somewhat dated user interface.
The program sometimes crashes; as no scripting interface is available, it can be difficult to restart the analysis from the point where it was interrupted.

XPPAUTO is used through a graphical interface.
Models are specified symbolically in text files, meaning no knowledge of coding is required.
Most features of AUTO are accessible, allowing users to exploit its powerful continuation routines without writing Fortran code.
XPP is capable of simulating many system classes, including ordinary, delay, and stochastic differential equations, boundary value problems, and difference and functional equations.
The package is written in C, and source code is released under the GNU GPL v2 license, allowing modification and redistribution.
Nevertheless, being a compiled GUI package, the code base does not easily lend itself towards being extended or adapted to novel problems.

XPPAUTO has a wide range of features: over a dozen solvers are available, covering forward and backward integration for a range of stiff and non-stiff classes of system.
Tools are also provided for phase plane analysis, and methods exist to create Poincar\'{e} sections and animations.

\subsection{PyDSTool}
\label{sec:org84712f2}

PyDSTool provides a suite of tools for the simulation and analysis of dynamical systems, with a focus on biological applications \cite{clewley2012hybrid}.
It is written primarily in Python3, however legacy C and Fortran code is included for efficient numerical solvers.
Being written in Python3, PyDSTool is particularly easy to adapt and extend to new problems.
The code is released under the permissive BSD 3-clause license, which allows modification and redistribution.
PyDSTool supports the simulation of ordinary differential equations, differential algebraic equations, discrete maps, and hybrid models thereof.
Limited support is also available for delay differential equations.

PyDSTool has no graphical user interface.
Instead, modelling and analysis procedures are specified through Python scripts.
This makes it easier to run and re-run complex analysis routines than with a graphical interface.
One can easily reproduce existing work, or change the model and analysis by re-running an altered script, without having to repeat the entire analysis from scratch.
Rich data structures are also provided, which can be easily integrated into other work.

As with XPP, PyDSTool specifies models symbolically.
Symbolic expression routines exist for manipulating derivatives, substitutions, evaluations, and for simplifying equations.
Continuation methods are implemented to detect and track bifurcations.
Toolboxes exist for a range of purposes, including parameter fitting and estimation, compartmental modelling of neurons and chemical synapses, and phase plane analysis.
These toolboxes are targeted towards computational biology.

PyDSTool is the only package to explicitly offer support for building and analysing hybrid models.
These can be loosely considered as a set of dynamical regimes, and a set of rules to dictate when and how to switch regimes; see \cite{simic2005towards} for a rigorous treatment of hybrid systems.
Hybrid modelling can allow one to express key system behaviours more simply than with smooth systems.
An example is the simple spike generation model in an integrate-and-fire neuron \cite{gerstner2014neuronal}, in which the hybrid model abstracts away the dynamics of spike generation, giving a simple model that captures the phenomenology of neuronal behaviours.

\subsection{\matcont{}}
\label{sec:orga52787b}

LOCBIF \cite{khibnik1992locbif} is an interactive bifurcation analysis tool, which has been superceeded by CONTENT \cite{kuznetsov1997content}; \matcont{} \cite{dhooge2008new}, in turn, emerged out of CONTENT.
\matcont{} focusses on the simulation and continuation of ordinary differential equations.
The package implements the detection, continuation, and normal form calculations of many bifurcations, and the intuitive graphical interface lends itself towards a relatively gentle learning curve.
\matcont{} is freely available under the Creative Commons BY-NC-SA 3.0 license, allowing users to modify and redistribute the software, subject to constraints.
Note that \matcont{} is written for use with MATLAB, and thus requires a MATLAB license.

\matcont{} is extensively documented, and a large number of tutorials are available.
It is available both as a graphical package, and as the command-line version \clmatcont{}.
This means that users are not required to write any code to use \matcont{}.
Nevertheless, the graphical interface acts only as an intermediary between the user and \clmatcont{}, so analyses can identically be carried out in a scripting environment.
\clmatcont{} also allows one to extend the functionality of the software, and integrate \matcont{} routines into custom projects.

\matcont{} has a built-in memory management system, where curves and points of interest are saved automatically.
As a result, users are not forced to re-run analyses from scratch each time a system is studied.

Models are written symbolically.
The software is able to compute symbolic derivatives, allowing for faster code execution and improved precision.
It supports additional features such as Poincar\'{e} maps and phase response curves, and is the only software to support normal form analysis of limit cycle bifurcations, using the methods developed in \cite{kuznetsov2005numerical}.
Users have access to all MATLAB solvers, in addition to two additional Runge-Kutta solvers for stiff systems.
Unlike with PyDSTool, no toolboxes are available for computational biology tasks.

\subsection{CoCo}
\label{sec:orgb7cb7d5}

CoCo (\emph{`Computational Continuation Core'}) is a software framework for running continuation experiments in MATLAB \cite{dankowicz2013recipes}.
The project is designed as a development platform for continuation toolboxes, from which application-specific routines can be constructed.
CoCo comes with the \emph{ep} toolbox, for continuation of equilibrium points; the \emph{coll} toolbox, for collocation-based discretisation of trajectory segments; and the \emph{po} toolbox, for continuation of periodic orbits.
These provide pre-built routines for studying a range of bifurcations.
`CoCo' is used here to refer to both the continuation core itself, and these bundled toolboxes.

Much like PyDSTool, CoCo is entirely scripting-based.
Models are specified as MATLAB functions, which are passed through various functions to set up and run a continuation experiment.
As with PyDSTool, the scripting design allows one to share the codes used to create any analyses.
Analyses can be run, modified, and re-run more easily than for graphical-only tools.
Nevertheless, the use of a scripting interface requires practitioners to be sufficiently competent with MATLAB programming.

CoCo has less support for detecting and continuing bifurcations than \matcont{} and PyDSTool; its capabilities are similar to those of XPPAUTO.
However, CoCo is designed specifically for extensibility -- users with knowledge of bifurcation analysis and programming will be able to design codes for handling arbitrary bifurcations.
With sufficient mathematical and coding skills, one can develop methods in CoCo to handle any problem that can be solved using continuation, making it a powerful tool for advanced users.

CoCo sits within a MATLAB workflow.
MATLAB routines can be used to simulate a system, search for fixed points, generate phase plane diagrams, etc.
The results of these procedures can then be used to determine how continuation should be applied.
This differs from the other tools discussed here, in which simulation, analysis and continuation are all available within one single package.

\subsection{Other continuation tools}
\label{sec:orgf65dff9}

Numerous other continuation packages exist, both for ODEs and other classes of system. 
These are discussed here.

The various versions of AUTO \cite{doedel1981auto,doedel1997auto,doedel2007auto} can be used for continuation experiments.
Note that PyDSTool \cite{clewley2012hybrid} and XPPAUTO \cite{ermentrout2002simulating} both provide AUTO interfaces.

DsTool \cite{back1992dstool} aims to provide an interactive package for all computations regarding dynamical systems.
PyDSTool is designed as a modern replacement for DsTool \cite{clewley2012hybrid}.
LINBLF \cite{khibnik1990linlbf} and CANDYS/QA \cite{feudel1992candys} are other tools for the bifurcation analysis of ordinary differential equations.

The introduction of time delays into a system can give rise to complex dynamics.
Systems become infinite-dimensional, and must be given initial data over a time interval.
Delay differential equations describe a range of phenomena in biology and physiology, in areas such as immunology \cite{marchuk2013mathematical}, neuronal interactions \cite{coleman1976theory}, and biochemical reactions \cite{macdonald1977time}; reviews are provided in \cite{an1979delays,bocharov2000numerical}.
Due to the difficulty in analysing delay equations, numerics are regularly used.
Two main packages exist for these purposes.
DDE BIFTOOL \cite{engelborghs2002numerical} is a scripting-based MATLAB package for analysis of systems with fixed delays.
It provides stability analysis and tracking of equilibrium and limit cycle solutions, and is capable of tracking bifurcations.
Knut \cite{szalaiknut} is a graphical package for analysing and simulating DDEs.
It supports stability analysis, orbit continuation and bifurcation detection in one parameter, and has methods for the two-parameter continuation of some bifurcations.
Unlike DDE BIFTOOL, it requires no programming knowlege to use; being written in C++, it is also faster.

Partial differential equations describe derivative problems over multiple independent variables.
This is often used to describe continuum problems in biology \cite{murray2001mathematical}.
PDEs have been used in developmental biology to model self-organisation and the emergence of complex structures in multicellular systems \cite{baker2008partial}, in ecology to model the spatial dynamics of populations \cite{holmes1994partial}, and in systems biology to describe the process of morphogenesis \cite{savill1997modelling}; see \cite{friedman2012pde} for a review.

PDECONT is a C library designed for continuation and bifurcation analysis of large time-evolving maps, representing discretised partial differential equations.
PDECONT is developed in \cite{roose1995newton,lust1998adaptive,lust2000computation}.
Users must have some degree of programming experience to use PDECONT, as the software comes as a library, rather than an interactive environment.
Similarly, LOCA \cite{salinger2002loca} provides a set of C libraries that can be used for continuation problems arising from high-dimensional systems, such as the discretisation of PDEs.
pde2path is a Matlab continuation and bifurcation analysis package for elliptic PDEs, in 1, 2, and 3 dimensions, over a range of geometries and boundary conditions \cite{uecker2014pde2path}.
The use of the high-level MATLAB language makes it easier to use than PDECONT.
WAVETRAIN uses continuation to discover regions of parameter space in which travelling waves exist in the solutions of PDEs, and to investigate the stability of those solutions \cite{sherratt2012numerical}.
The software uses text-based input files, meaning no coding is required.

Nonsmooth systems can be artificially created to simplify the study of naturally occuring smooth systems \cite{jeffrey2018hidden}.
The quintessential example of this is the integrate-and-fire neuron, whereby piecewise linear dynamics can be used to simplify the study of large networks of spiking neurons \cite{coombes2012nonsmooth}.
These simple nonsmooth neuron models are the model of choice for simulating large, complex networks, owing to their biophysical realism, and the low computational demands for simulating them \cite{izhikevich2003simple}.
Two main packages exist for the continuation of bifurcations in nonsmooth systems.
SlideCont is an AUTO97 \cite{doedel1997auto} driver for studying sliding bifurcations in discontinuous piecewise-smooth systems \cite{dercole2005slidecont}.
Hybrid systems, as discussed in section \ref{sec:org84712f2}, are a special type of nonsmooth system; PyDSTool \cite{clewley2012hybrid} and TC-HAT \cite{thota2008tc} use numerical continuation to track periodic trajectories in hybrid systems. 

Maps are models where state variables are iteratively updated.
Maps often arise in biology as discrete-time models of population dynamics \cite{murray2007mathematical}.
They can also be derived from continuous-time flows, in the form of Poincar\'{e} maps \cite{strogatz2018nonlinear}.
The most powerful map continuation tool is \matcontm{} \cite{govaerts2008cl}; it is also the only tool designed specifically for continuation problems on maps.
Nevertheless, several of the ODE packages also provide continuation methods for maps, including CONTENT \cite{kuznetsov1997content}, DSTool \cite{back1992dstool}, and PyDSTool \cite{clewley2012hybrid}.

\section{Conclusion}
\label{sec:org6aa7924}
Here the role of bifurcation analysis in mathematical biology has been illustrated.
Key ideas from nonlinear dynamics and bifurcation theory were introduced.
Examples have been provided to highlight the uses of numerical continuation, and a case study has been used to demonstrate how to run a numerical continuation experiment; this was used to explore the causes of bursting behaviours in a simple neuron model.
Software tools have been discussed for studying bifurcations in a range of systems.
The capabilities of the four most commonly used ODE tools were discussed in depth.
The author's usage suggestions are given here, based on these capabilities.

For practitioners with minimal coding experience, XPPAUTO and \matcont{} are recommended, owing to their graphical user interfaces.
Those with coding experience are recommended to try PyDSTool or CoCo; the use of scripts makes analyses repeatable, easier to collaborate on, and easier to adapt.
Where speed is the primary concern, XPPAUTO and PyDSTool are the most appropriate choice, as a result of their highly optimised numerical algorithms.
When complex bifurcation structures are present, \matcont{} is recommended, as it is capable of detecting and tracking the most types of bifurcations.
For extending and building the tools into more sophisticated continuation procedures, CoCo and PyDSTool are recommended, as their scripting interface allows for easy extensions; CoCo is designed as a development platform, and is thus particularly suited to this task.
When analyses are conducted beyond bifurcations, XPPAUTO and PyDSTool excel.
They both offer a variety of tools for phase plane diagrams, including nullcline and flow field plotting, and equilibrium searching.
It is also noted that PyDSTool comes with toolboxes for a wide range of computational biology problems; PyDSTool is recommended where relevant toolboxes are available, as they can greatly simplify system analysis.

\section{Acknowledgements}
\label{sec:org007eb7f}
M.B. was supported by an EPSRC DTP Scholarship, provided by the University of Bristol.
L.M. was funded by the Engineering and Physical Sciences Research Council (EPSRC, grants EP/R041695/1 and EP/S01876X/1) and Horizon 2020 (CosyBio, grant agreement 766840).
L.R. has received funding from the Royal Academy of Engineering (RF1516/15/11) which is gratefully acknowledged.

\section{Conflict of interest}
\label{sec:org01b83eb}
The authors declare that they have no conflict of interest.

\section{Author contributions}
\label{sec:org8776716}
M.B. wrote the manuscript. L.M and L.R conceived of the review topic, and provided guidance and feedback in the writing of the manuscript.

\bibliographystyle{unsrt}
\bibliography{references}
\vspace{2in}
\end{document}